\documentclass{osa-article}

\journal{osajournal}

\usepackage{graphicx}
\articletype{Research Article}

\begin{document}

\title{Raising the stimulated Brillouin scattering threshold power by longitudinal compression gradient in a fiber amplifier}

\author{Anasthase Liméry\authormark{*}, François Gustave, Laurent Lombard, Anne Durécu, and Julien Le Gouët\authormark{\dag}}

\address{Office National d'Etudes et de Recherches Aérospatiales, Palaiseau, France}
\email{\authormark{*}anasthase.limery@onera.fr} 
\email{\authormark{\dag}julien.le\_gouet@onera.fr} 

\begin{abstract}
We demonstrate and analyse a novel approach to enhance the threshold power of stimulated Brillouin scattering (SBS) in optical fibers, using a longitudinal compressive strain gradient. We derive analytical expressions for the power spectral density of the backscattered Stokes wave in the general case of passive and amplifying optical fibers, by considering the strain and optical power distributions. Our method provides an accurate prediction of the SBS gain spectrum, that we illustrate with a quantitative comparison between measurements and calculations of the SBS Stokes spectra, before and after applying the compression gradient. Our experimental results demonstrate the successful enhancement of the SBS threshold power by a factor of about 3 for the passive fiber and 2 for the amplifying fiber. The enhancement that we manage to calculate in the case of the passive fiber is in perfect agreement with the experimental result.
\end{abstract}

\section{Introduction} 

Numerous transportable laser systems are based on all-fiber sources, which offer compactness and low sensitivity to vibrations. Applications like long range wind velocity measurements require the emission of high peak power laser pulses with narrow spectral linewidth. However, due to the long interaction length and high confinement of the optical field, nonlinear effects can bridle the output power of fiber lasers. In particular, high power coherent fields with spectral linewidth narrower than a few MHz can be scattered back almost entirely by Stimulated Brillouin Scattering (SBS) \cite{Kobyakov_2010}.

The effect of SBS can be circumvented, to raise the power of single-frequency sources, with several methods: increase the effective mode area, or reduce the interaction length, or reduce the SBS gain \cite{Cavillon_2018}. One way to reduce the SBS gain consists in applying a gradient of the glass density along the fiber, as detailed in the next section. In practice, this can be obtained through thermal or stress gradients. 

For optical fibers with a standard acrylate coating, the thermal gradient is limited by the range of temperature that this polymer can endure safely for long term operation. Typically the thermal amplitude rarely exceeds 80°C, which barely allows a twofold increase of the maximum power \cite{Imai_1993, Lou_2020}.

The other approach to modify the glass density, by mechanical constraints, was also early identified \cite{Horiguchi_1989}. Soon after, the principle of an inhomogeneous strain distribution to reduce the effect of SBS were demonstrated and analyzed \cite{Yoshizawa_1993, Shiraki_1996}. Since then, many works have been dedicated to study this effect and increase the SBS critical power \cite{Boggio_2005, Mamdem_2012, Zhang_Yb_2013, Lucas_2014, Engelbrecht_2014}. Yet for most of the methods reported to date, the spectral broadening was obtained by a tensile strain. This is obviously the simplest way to apply a mechanical constraint on a thin fiber, but the lifetime of the fiber under tension is a sensible concern. 
On the other hand, fused silica can resist much higher compression than tension. More precisely, the compressive strength of silica is about 20 times higher than the tensile strength. 

Few methods of compression have been proposed in the past: one was based on a coiling arrangement \cite{Rothenberg_2008}, which reduces the interest of optical fibers for the transport of light; another one consisted in stranding fibers around a wire, and stranding again several such ensembles around a cable \cite{Yoshizawa_1993}. Although this configuration demonstrated an impressive broadening of the SBS gain linewidth from 50~MHz to 400~MHz on a 17~km-long fiber, no report was published for its application to a fiber amplifier. 

In this paper we detail the first demonstration of a compressive stress applied on straight passive and active optical fibers. Section \ref{sec_Principle} is a brief description of the numerical model that we use to predict the SBS gain broadening, for a given stress profile. The fabrication process is described in section \ref{sec_Fabrication}, and we present and discuss the experimental results in section \ref{sec_Passsive_fiber} for passive fibers, and section \ref{sec_Active_fiber} for amplifying fibers.

\section{Physical principle and analytical modeling}
\label{sec_Principle}

The generation of Brillouin scattering relies on a phase-matching condition between the optical fields (pump and Stokes), and the acoustic waves \cite{Kobyakov_2010}. The momentum of the latter is determined by the local acoustic velocity, which is related to the material density. Therefore, dividing the effective fiber length into several segments with different material densities prevents the phase-matching from building along the whole fiber length. In particular, it is well-known that a strain gradient can induce the SBS gain inhomogeneous broadening that raises the SBS threshold power $P_\text{th}^\text{SBS}$ \cite{Yoshizawa_1993}.

The originality of our technique consists in applying a compressive gradient along a straight (uncoiled) fiber, contrary to the more usual tensile strain \cite{Shiraki_1996, Boggio_2005}. However, the calculations of the spectra of SBS gain and Stokes power spectral density (PSD) are identical, so we will only briefly remind them. Thereafter, we refer to the single-frequency input optical power that generates the SBS as the SBS pump power $P_\text{p}(z)$ for a position $z$ along the fiber, and we define as $P_\text{S}(z)$ the power of the Stokes optical field that is backscattered by SBS effect. The PSD of the Stokes field is noted as $p_\text{S}(\nu)$ and at a given position, we have $P_\text{S}(z)=\int p_\text{S}(z,\nu) \,\mathrm{d}\nu$.

Since we are interested in developing methods to raise $P_\text{th}^\text{SBS}$ and avoid significant power backscattering, we consider the \textit{undepleted pump approximation} where $P_\text{p}$ is not significantly transferred to the Stokes wave. In a single-frequency fiber amplifier, the laser signal presents a long coherence time compared to the acoustic waves lifetime, hence playing the role of a pump for the SBS process. Therefore we represent the laser amplified signal power as $P_\text{p}(z)$ . Concerning the laser amplifier, the gain variation $g_a(z)$ along the fiber is obtained from the expression:
\begin{equation}
\label{eq_def_gain}
g_a(z)=\frac{1}{P_\text{p}(z)}.\frac{\mathrm{d}P_\text{p}(z)}{\mathrm{d}z}
\end{equation}

Let us consider that the Stokes field propagates in the positive direction ($\mathrm{d}z>0$), and the SBS pump (i.e. the "useful" laser signal) in the negative direction ($\mathrm{d}z<0$). For a scalar optical field (single polarization state and transverse mode), the variation of the Stokes PSD $p_\text{S}(\nu)$ is given by the following inhomogeneous linear ordinary differential equation \cite{Engelbrecht_2014}:
\begin{equation}
\label{eq_ED_DSP}
\dfrac{\mathrm{d} p_\text{S}}{\mathrm{d}z}(z,\nu)=g_a(z).p_\text{S}(z,\nu)+\gamma_\text{B}(z,\nu)p_\text{S}(z,\nu)P_\text{p}(z) + h\nu_\text{S} \left(1+N_\text{BE}\right) P_\text{p}(z)
\end{equation}
where $\nu_S$ is the optical frequency of the Stokes field, $\gamma_\text{B}(z,\nu)$ the SBS gain spectrum as a function of longitudinal position, $N_\text{BE}$ the Bose-Einstein occupation number for acoustic phonons of frequency $\nu_\text{ac}$. At room temperature $T_r$, we have $N_\text{BE}\simeq k_B T_r/h\nu_\text{ac}\gg 1$. The first term on the right-hand side describes the longitudinal amplification or attenuation of the Stokes wave, the second term corresponds to its SBS amplification, and the last term is the contribution of the SBS amplified spontaneous emission.

The spectral/spatial map of the SBS gain $\gamma_\text{B}(z,\nu)$ is simply inferred from the strain profile $\epsilon(z)$ that is applied on the fibers. Indeed for elastic stress, it is well-known that the frequency of the Stokes wave $\nu_S$ varies linearly with strain \cite{Horiguchi_1989}. For an arbitrary longitudinal variation of the strain $\epsilon(z)$, 
the evolution of the SBS gain as a function of the position writes as:
\begin{equation}
\label{eq_SBS_gain_spatial}
\gamma_\text{B}(z,\nu)= \dfrac{\gamma_0}{1+\left(2\cdot\dfrac{\nu-\nu_S-C_\epsilon\epsilon(z)}{\Delta\nu_\text{SBS}}\right)^2}
\end{equation}
where $C_\epsilon$ is the coefficient of spectral sensitivity to the strain (in MHz/\% of length variation), which depends on the glass composition.

We first consider the case of a short passive fiber, without pump amplification or significant attenuation ($g_a=0$), so $P_\text{p}$ is constant along the fiber. The Stokes PSD can be obtained simply by integrating $p_\text{S}(z,\nu) + h\nu N_\text{BE}$. Throughout the paper, we follow the choice of referential from \cite{Engelbrecht_2014} and consider that the positive propagation direction (d$z>0$) is that of the SBS Stokes wave, so that \textit{the SBS pump wave propagates in the negative direction}. The initial condition for the Stokes integration is $p_\text{S}(z,\nu)=0$ at the fiber input $z=0$. We thus find at the fiber output ($z=L$):

\begin{equation}
\label{eq_DSP_passive}
p_\text{S}(\nu)=h\nu_S N_\text{BE}\left[ \exp\left(P_\text{p}\int_0^L \gamma_\text{B}(z,\nu)\,\mathrm{d}z \right) -1\right]
\end{equation}



As soon as the power variations of the pump and Stokes fields become significant along the propagation length, the longitudinal evolution $g_a(z)$ of the optical powers must be taken into account. In that case the equation \ref{eq_ED_DSP} is first solved without the last term, yielding an exponential solution with a constant prefactor. Then we solve the complete equation by letting the prefactor vary, which yields the following general solution :

\begin{equation}
\label{eq_DSP_active}
p_\text{S}(\nu)=h\nu_S N_\text{BE} \int_0^L \gamma_\text{B}(\nu,z) P_\text{p}(z) \exp \left( \int_z^L \left[ g_a(z') +\gamma_\text{B}(\nu,z')P_\text{p}(z') \right] \mathrm{d}z' \right)
\mathrm{d}z
\end{equation}

It should be noted that a similar expression was presented in \cite{Engelbrecht_2016}, which is only valid when the local gain $g_a$ (or $g$ in the article) does not depend on the position $z$.

In the general case of a fiber amplifier, the optical gain experienced by the laser field is not constant. This laser signal corresponds to the SBS pump, and its gain $g_a(z)$ can be derived from a separate numerical simulation. Due to the choice of propagation direction, in the negative $z$ direction for the SBS pump wave (i.e. the laser amplified signal), the amplifier gain $g_a(z)$ in equation \ref{eq_DSP_active} corresponds to the reversed gain $g_a(-z)$ extracted from the simulation of the laser amplification.

The comparison between these calculations and the experimental results will be discussed in the next sections (\ref{sec_Passsive_fiber} for passive fibers and \ref{sec_Active_fiber} in the case of an Er-doped fiber), for the typical compressive strain profiles $\epsilon(z)$ that we manage to obtain. 



\section{Compression process}
\label{sec_Fabrication}

By principle, an optical fiber exhibits a giant aspect ratio, as defined by the ratio between the length and the diameter. The main difficulty to apply a longitudinal compressive strain on a fiber section object is to avoid any micro-bending that could damage the material, or most generally induce high optical loss. After working on a technique of progressive tensile strain gradients \cite{Canat_2016}, we developed a process to compress optical fibers without noticeable micro-bending \cite{Canat_CLEO_2016}.

\begin{figure}[ht!]
\centering\includegraphics[width=9cm]{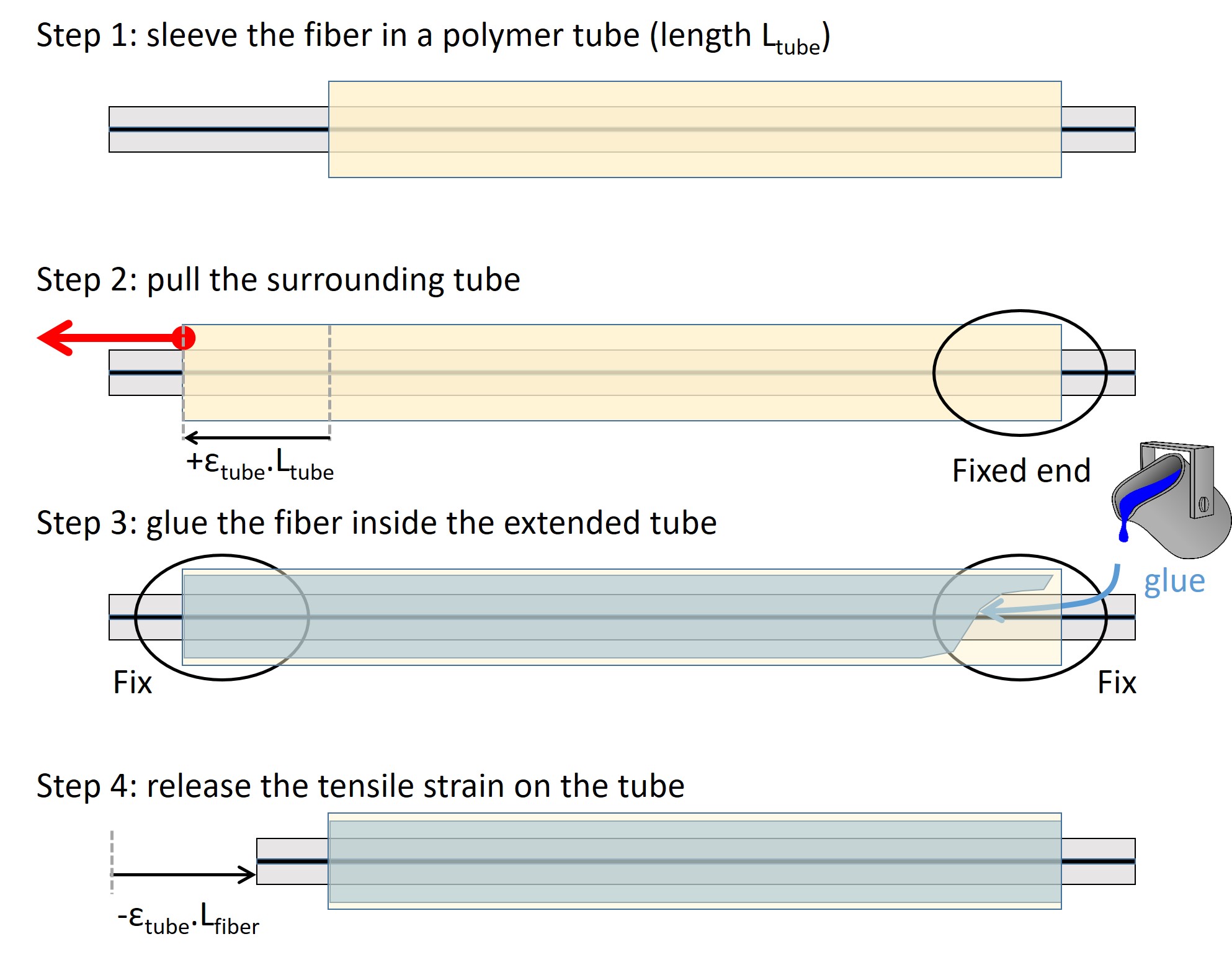}
\caption{Description of the fabrication process to apply a permanent compressive stress on one fiber segment.}
\label{fig_strain_method}
\end{figure}

The method to apply a compressive strain on a fiber follows 4 main steps depicted in Figure~\ref{fig_strain_method}. Firstly, the fiber is inserted into a flexible tube, which extremities are fixed. Then this tube is elongated thanks to a translation stage placed at one of its extremities. In the third step, a glue is introduced inside the tube and the fiber is bonded inside the elongated tube. Finally, the tube is released and tends to move back to its initial length. The glue hinders the tube releasing and transmits the longitudinal compression to the fiber which is now integral to the tube. The resulting frequency shift of the Brillouin spectrum is controlled thanks to initial tube elongation.

To broaden the SBS gain spectrum of the fiber, every section of the fiber should undergo a different strain. This is achieved by dividing fictitiously the fiber in N sections, and repeating N times the steps 3 and 4. For each repetition, only one section is bonded (which is possible thanks to a UV-curable glue) and the tube is released gradually step by step.

The main risk in this technique is inducing losses in the fiber because of buckling. To avoid this failure, the tube and the glue must be carefully chosen while taking into account the characteristics of the fiber (diameter and Young modulus $E_1$ of the silica). On one side, the tube should present a high elasticity and a high Young modulus $E_3$. On the other side, the glue should be curable and present a sufficiently high Young Modulus $E_2$ (at least 200~MPa). The inequality $E_1.S_1 + E_2.S_2 < E_3.S_3$ must be satisfied, where $S_1$, $S_2$ and $S_3$ are respectively the sections of the fiber, the glue and the tube, so that the compression could be successfully transmitted to the fiber without fiber buckling or tube distortion. Let us note that the final packaging has a very low impact on the handling of the optical fiber.



\section{Demonstration on a passive fiber}
\label{sec_Passsive_fiber}

We first apply our method on a 1~m single-mode PM fiber (Corning PMF1550), intended as the output delivery pigtail of a pulse fiber amplifier for a lidar system. After measuring the compression profile, we calculate the expected PSD of the SBS Stokes field (eq. \ref{eq_DSP_passive}), and finally compare the calculated spectrum to the direct measurement of the Stokes PSD. The compression is applied over a 0.7~m portion, on 7 incremental steps, each about 10~cm long. 

\subsection{Calculation of the Stokes PSD from the strain profile measurement}
\begin{figure}[ht!]
\centering\includegraphics[width=9.5cm, trim={0 .5cm 0 0},clip]{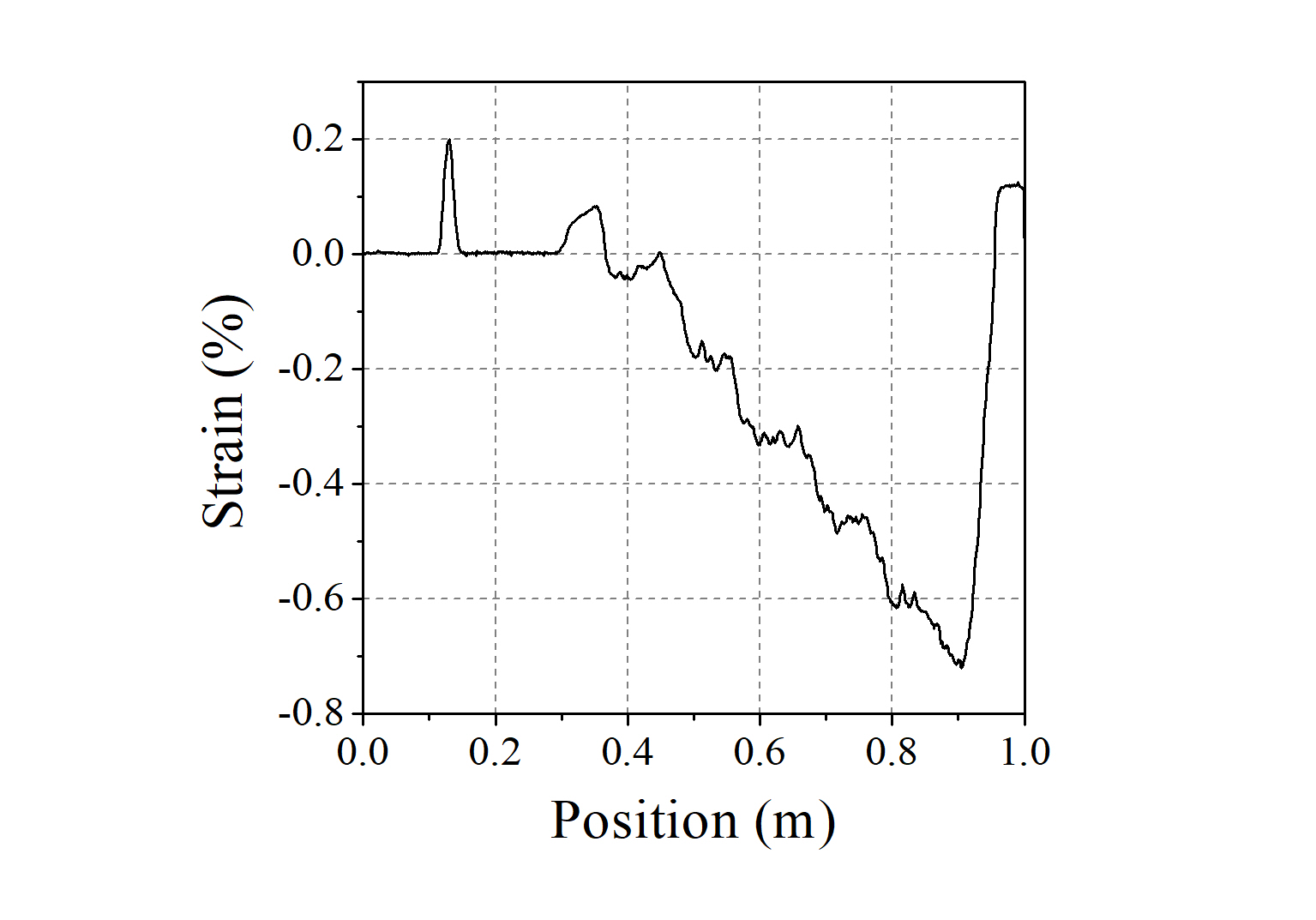}
\caption{Compressive strain profile of the passive fiber that leads to the calculation of the SBS Stokes spectrum in figure \ref{fig_DSP_passive_SMF} (right).}
\label{fig_strain_profile}
\end{figure}

The strain profile $\epsilon(z)$ along the fiber is measured by probing the fiber with an optical frequency domain reflectometry (OFDR) commercial system (OBR 4613, from Luna Innovations), before and after the compression process. The instrument uses a tunable laser source to measure the Rayleigh backscattering spectra along the fiber, in the two configurations. By calculating the cross-correlation between the two spectra (without and with strain), it can provide the strain profile $\epsilon(z)$, as reported on Figure \ref{fig_strain_profile}, for a spatial resolution of about 2~mm. The spectral/spatial map of the SBS gain $\gamma_\text{B}(z,\nu)$ is then obtained from equation \ref{eq_SBS_gain_spatial}, with $C_\epsilon = 430$~MHz/\%.


The true SBS Stokes PSD is measured by collecting on a fast photodetector (>12~GHz bandwidth) the interference between a local oscillator and the Stokes field generated by a pump field in the Fiber Under Test (FUT) \cite{Yeniay_2002}. In practice, we measure two SBS Stokes spectra : the raw spectrum contains the contribution of the whole setup (fiber components and FUT), and a "reference" spectrum is measured by blocking the signal propagation at the beginning of the FUT. The fiber path difference between the two configurations corresponds to the FUT length, here very close to 1~m. The Stokes PSD is then obtained by subtracting the two spectra.

\begin{figure}[ht!]
\centering\includegraphics[width=\linewidth,trim={.5cm 0 0 0},clip]{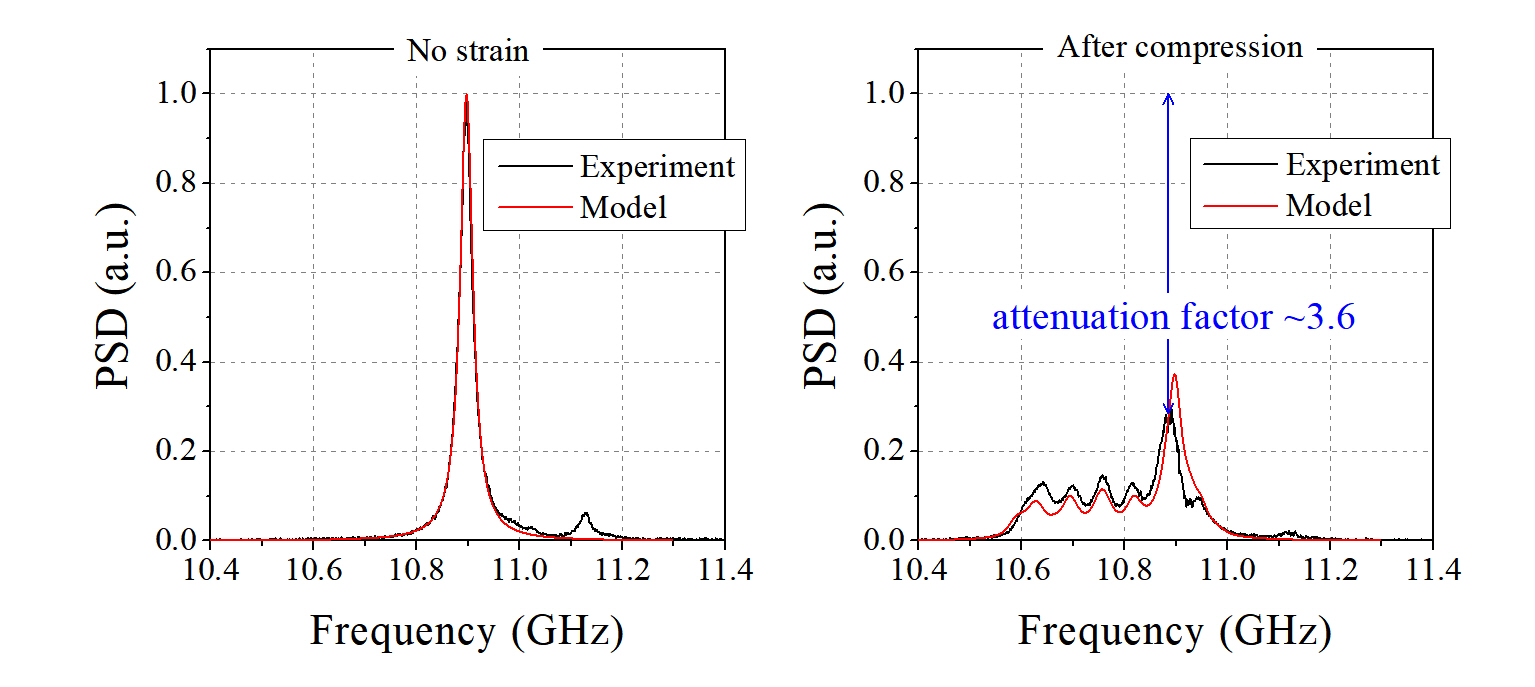}
\caption{Power spectral densities of the SBS Stokes waves measured (black) and calculated (red) on a PMF1550 fiber, before (left) and after (right) applying the compressive strain gradient illustrated in Fig. \ref{fig_strain_profile} along the fiber.}
\label{fig_DSP_passive_SMF}
\end{figure}

The measurements before and after compression process are gathered on Figure \ref{fig_DSP_passive_SMF} (respectively left and right), together with the calculated Stokes PSD $p_\text{S}(\nu)$, derived with the model presented above (see eq. \ref{eq_DSP_passive}). The SBS spectrum width $\Delta\nu_\text{SBS}$ of the model has been set to fit optimally the experimental curve without strain with a value of 30~MHz, in good agreement with the usual values reported in silica fibers \cite{Kobyakov_2010}. The amplitudes of the four curves are normalized to the maximum of the spectrum obtained without strain.
We find here a good agreement between calculation and measurement of the SBS Stokes PSD for the strained fiber, both in amplitude and shape. More precisely, the measurement shows an attenuation $a_\text{meas}$ of the peak Stokes PSD by a factor $a_\text{meas} \simeq 3.6$, whereas the calculation predicts an attenuation $a_\text{calc}=2.7$. 

\subsection{Estimation of the SBS threshold power improvement factor}
The next step consists in using the calculated Stokes PSD to infer the influence of the spectral broadening on the SBS threshold power $P_\text{th}$. We quantify here this evolution by the Power Improvement Factor (PIF), defined as the ratio  $\text{PIF}=P_\text{th}^\text{after}/P_\text{th}^\text{before}$ between the SBS threshold powers after and before applying the strain.

The SBS threshold power can be defined as the input 'pump' power $P_\text{P}(z=L)$ for which the output power $P_\text{S}(z=L)$ of the SBS Stokes wave equals, at the end of its propagation, a small fraction $\mu$ of $P_\text{P}(z=0)$ \cite{Engelbrecht_2014}. The parameter $\mu$ can be seen as the optical reflectivity caused by SBS. As for the SBS Stokes power, it is generally measured on a photodetector without spectral resolution, so it corresponds to the spectral integration of the power spectral density: $P_\text{S}(z=L)=\int p_\text{S}(\nu) \,\mathrm{d}\nu$.

Below the SBS threshold, it can be shown analytically that the SBS Stokes power does not depend on the spectral broadening, so $P_\text{S}^\text{after}=P_\text{S}^\text{before}$. Reaching higher input pump power (\textit{stimulated} Brillouin scattering), methods to broaden the SBS Stokes spectrum can slow down the increase of $P_\text{S}(z=L)$ and the reflectivity $\mu$. Using eq. \ref{eq_DSP_passive}, we can calculate the evolution of $P_\text{S}(z=0)$ as a function of the input power $P_\text{P}$, and thus determine for an arbitrary value of $\mu$ the corresponding value of $P_\text{P}$, before and after the strain gradient. In the case of the passive fiber described above, we obtain a reflectivity $\mu=1\%$ for 33~W before strain, but 107~W once the strain gradient on Fig. \ref{fig_strain_profile} is taken into account (considering $g_\text{B}=1\,10^{-11}$~m/W). Therefore, the calculation based on the strain measurement predicts an improvement factor PIF=3.2.

\subsection{Measurement of the SBS threshold power, before and after strain}
To test the relevance of this calculation, we proceed to the direct measurement of the SBS threshold power in the FUT, before and after compression. Since there is no exact or universal definition of $P_\text{th}$, several experimental methods are possible. In a previous study, we found that even the most empirical method, which consists in detecting an arbitrary alteration of the output pulse shape, yields the same relative variations of $P_\text{th}$ \cite{Le_Gouet_2019}. We thus use this method to determine the increase of the SBS critical power.

A source of relatively high peak power single-frequency laser pulses is necessary to probe the influence of the strain on the SBS critical power. We thus realize a pulse laser amplifier using the same Er:Yb doped fiber as described in the next section, delivering 200~ns laser pulses at a repetition rate of 20~kHz with a spectral linewidth of about 2~MHz. Thanks to a similar compressive gradient as described thereafter, the pulse peak power can reach almost 300~W peak power. The passive PMF1550 sample is then spliced to the output of the fiber amplifier. The peak power of the probe pulses is tuned via the pump power, while maintaining the pulse duration and repetition rate constant. The SBS critical power is about 70~W on the unstrained fiber, and reaches 220~W once the compression gradient has been applied. This evolution represents a threshold improvement by a factor PIF=3.1, in good agreement with the calculation using the actual strain profile, which predicted a factor of 3.2. Both figures are also close to the attenuation of the Stokes spectrum peak, measured as $a_\text{meas} \simeq 3.6$ (see Fig. \ref{fig_DSP_passive_SMF}).


\subsection{On the difference between PIF and Stokes spectrum variations}
\label{sec_PIF}
We must emphasize here that the similarity between the PIF and the attenuation $a$ of the Stokes spectrum peak is quite accidental. As an analytical example, let us consider a linear strain gradient with an amplitude that results in a nearly flat Stokes spectrum with a 3~dB bandwidth $\Delta \nu_\text{strain}=450$~MHz. With respect to the initial Lorentz profile with linewidth $\Delta \nu_\text{SBS}=30$~MHz, the spectrum is thus broadened by a factor of 15. For an input power below the SBS threshold, the reduction of the peak amplitude can be calculated analytically and one can show that it is equal to $a=2\Delta \nu_\text{strain}/\pi\Delta \nu_\text{SBS}\simeq10$. However, applying the calculations above to compare the SBS threshold powers yields a theoretical factor $\text{PIF}\simeq6$, so in this case, there is no identity between the broadening factor, the PIF, or the reduction of the Stokes spectrum amplitude. Eventually, this example shows that a detailed calculation is necessary in general to estimate the PIF from the spectral broadening.

\section{Case of a rare-earth doped amplifying fiber}
\label{sec_Active_fiber}

We then aim at applying the same method and analysis to a rare-earth doped double-clad fiber. The objective here is to improve the maximum power that can be delivered by a single-frequency laser amplifier (MOPA). 

\subsection{Calculation and measurement of the SBS Stokes wave PSD}
Our test fiber is an erbium-ytterbium co-doped fiber manufactured by iXblue Photonics (reference IXF-2CF-EY-PM-12-130-0.21), with a double-clad to allow optical pumping with multimode laser diodes. The core diameter and numerical aperture (NA) are 11~µm and 0.2, with a core absorption of 50~dB/m at 1535~nm by erbium ions. The clad diameter and NA are 124~µm and 0.46, with a clad absorption of 2.85~dB/m at 915~nm. This value corresponds to a core absorption by the ytterbium ions of approximately 1000~dB/m at 976~nm. The fiber amplifier is seeded by nearly Gaussian signal pulses with a pulse duration of 400~ns, a repetition rate of 30~kHz, and a peak power of 5~W. The pump energy is provided by a fiber-coupled 50~W multimode laser diode at 976~nm, coupled into the active fiber clad, co-propagating with the laser signal, through a fiber pump combiner.

For this configuration, we find that the optimal length of the active fiber is close to 2.6~m. 
Since the SBS Stokes power increases exponentially with the SBS pump power (here the signal pulses), we apply the strain at the end of the fiber over 1.2~m : the linear gradient consists of 12 sections, about 10~cm each. At the active fiber output, we splice a passive pigtail of Corning PMF to allow easy connection to the lidar components.

\begin{figure}[ht!]
\centering\includegraphics[width=10cm]{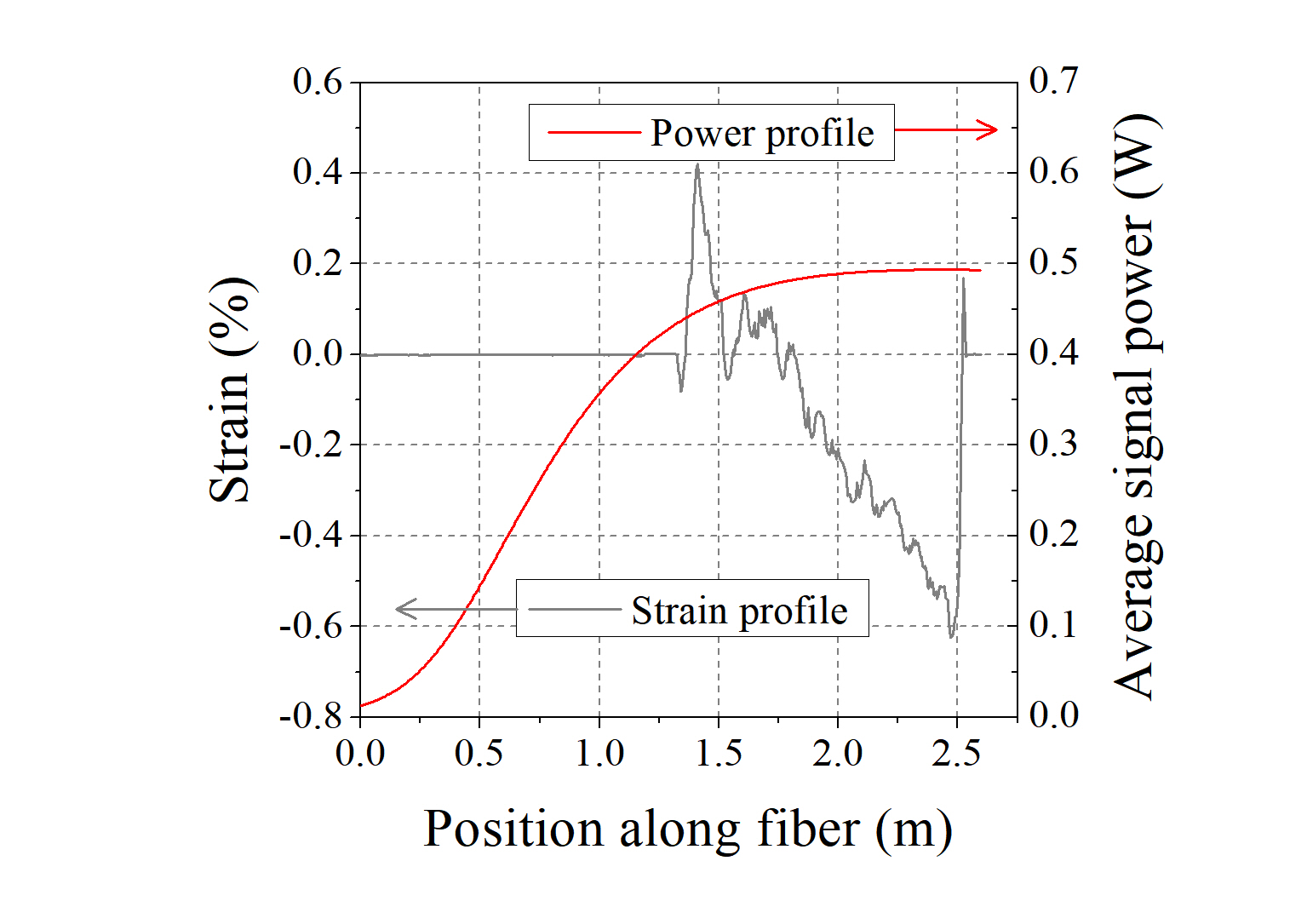}
\caption{Profiles along the amplifying fiber for the strain measured with the OFDR instrument (grey curve, left scale), and for the laser signal power calculated with the ONERA numerical simulation (red curve, right scale).}
\label{fig_MOPA_strain_profile}
\end{figure}

The strain profile $\epsilon(z)$ required in eq. \ref{eq_DSP_active} is measured with the same OBR~4613 instrument, and reported on Figure \ref{fig_MOPA_strain_profile} (grey curve). The wavelength domain of the probe laser is centered at about 1.3~µm, in order to avoid ground state absorption by erbium or other usual rare-earth ions (ytterbium, thulium, holmium...).

The other profile required to use eq. \ref{eq_DSP_active}, and predict the spectrum of the SBS Stokes field, is that of the amplifier gain $g_a(z)$ along the fiber. Once we realize the amplifier with $L_\text{FUT}=2.6~$m, we measure the signal power as a function of the pump power. Then we use our numerical simulation of fiber laser amplification, considering the fiber dimensions listed above, and the absorption and emission cross-section spectra reported in \cite{Le_Gouet_2019} for Er$^{3+}$ in our alumino-phospho-silica host. We adjust some parameters of the MOPA simulation, in particular the Yb-Er energy transfer $K_\text{tr}$, to fit the signal measurements (dashed curve on fig. \ref{fig_MOPA_SBS_treshold}), and infer the longitudinal profile of the gain $g_a(z)$ from the signal power evolution $P_\text{p}$ (Figure \ref{fig_MOPA_strain_profile}, red curve). In the MOPA simulation, we assume that the strain has no noticeable influence on the interaction between pump, signal and gain medium.

\begin{figure}[ht!]
\centering\includegraphics[width=1.0\linewidth,trim={.5cm 0cm 0 0},clip]{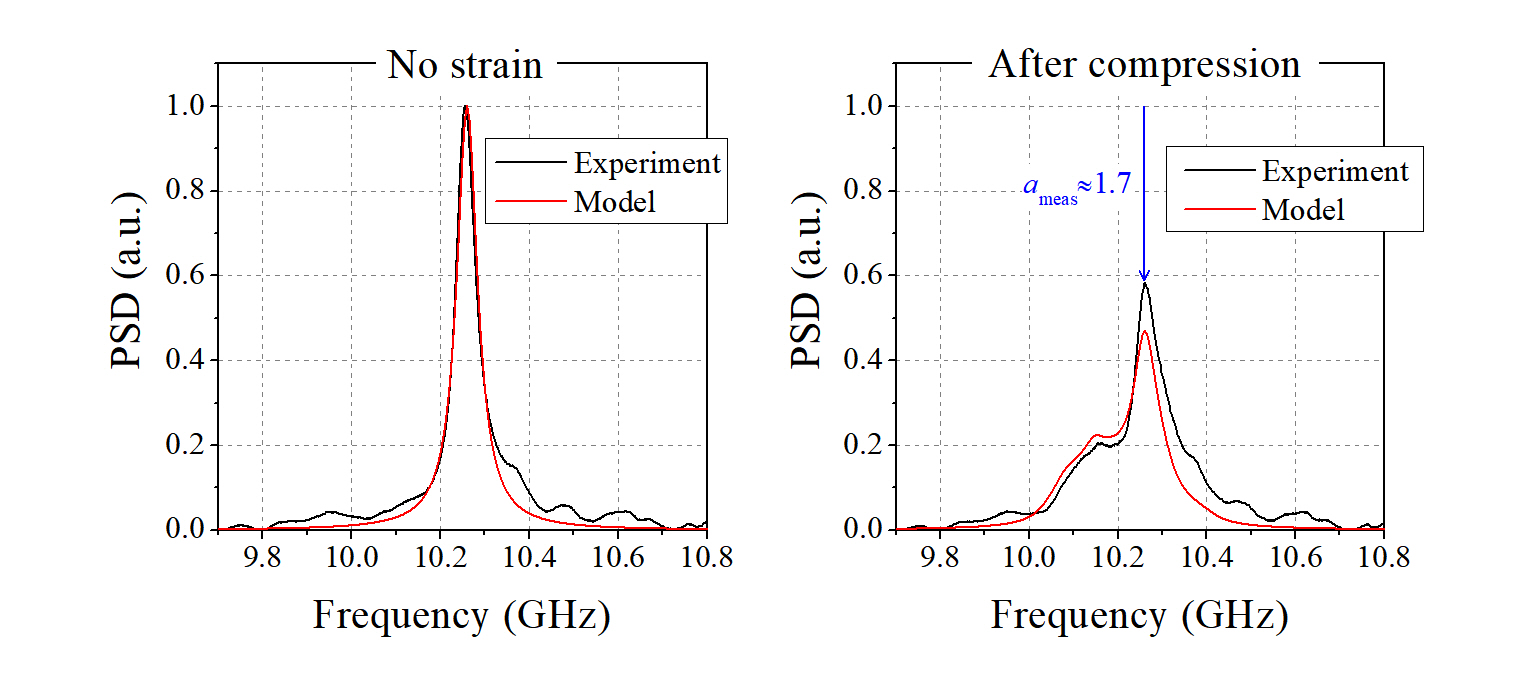}
\caption{Power spectral densities of the SBS Stokes waves measured (black) and calculated (red) at the active fiber input, before (left) and after (right) applying the compressive strain gradient illustrated in Fig. \ref{fig_MOPA_strain_profile} (gray line) along the fiber.}
\label{fig_DSP_MOPA_AL}
\end{figure}

Knowing the gain and strain profiles along the fiber, we can then calculate the spectrum of the Stokes field amplified by the combined gains of SBS and erbium ions (eq. \ref{eq_DSP_active}). The result of this calculation is illustrated on Figure \ref{fig_DSP_MOPA_AL} for the active fiber with and without strain, together with the results of the measurements in the same configurations. For the sake of the components of the SBS spectral analysis setup, the Stokes PSD are measured with the signal power limited to 500~mW at the amplifier output (i.e. a 976~nm pump power of about 5.4~W).

Concerning the measurement of the Stokes PSD in the MOPA configuration, the contributions of the passive components are more difficult to correct than in the previous configuration. Since the laser source was needed for lidar tests, we could not cut the active fiber to remove its contribution. Therefore we simply subtract a linear baseline that is common to the spectra obtained with or without strain. In order to fit optimally the experimental spectrum for the MOPA without strain (a typical lorentzian shape), the SBS spectrum width $\Delta\nu_\text{SBS}$ of the model has been set to 60~MHz, again in agreement with the common values for alumino-silicate fibers \cite{Dragic_2012}. Again the amplitudes of the four curves presented on Fig. \ref{fig_DSP_MOPA_AL} are normalized to the maximum of the measured PSD without strain. 

After compression of the MOPA fiber, the Stokes PSD maximum decreases by a factor $a_\text{meas}=1.7$ from its initial value. The calculation result is quite similar in shape and amplitude, predicting a reduction by a factor $a_\text{calc}=2.1$. The main cause of the discrepancy resides probably in the difference between the actual local gain $g_a(z)$ and the gain profile that we obtain from the MOPA numerical simulation.

\subsection{Demonstration of the SBS threshold increase}

\begin{figure}[ht!]
\centering\includegraphics[width=.7\linewidth,trim={0cm .5cm 0 0},clip]{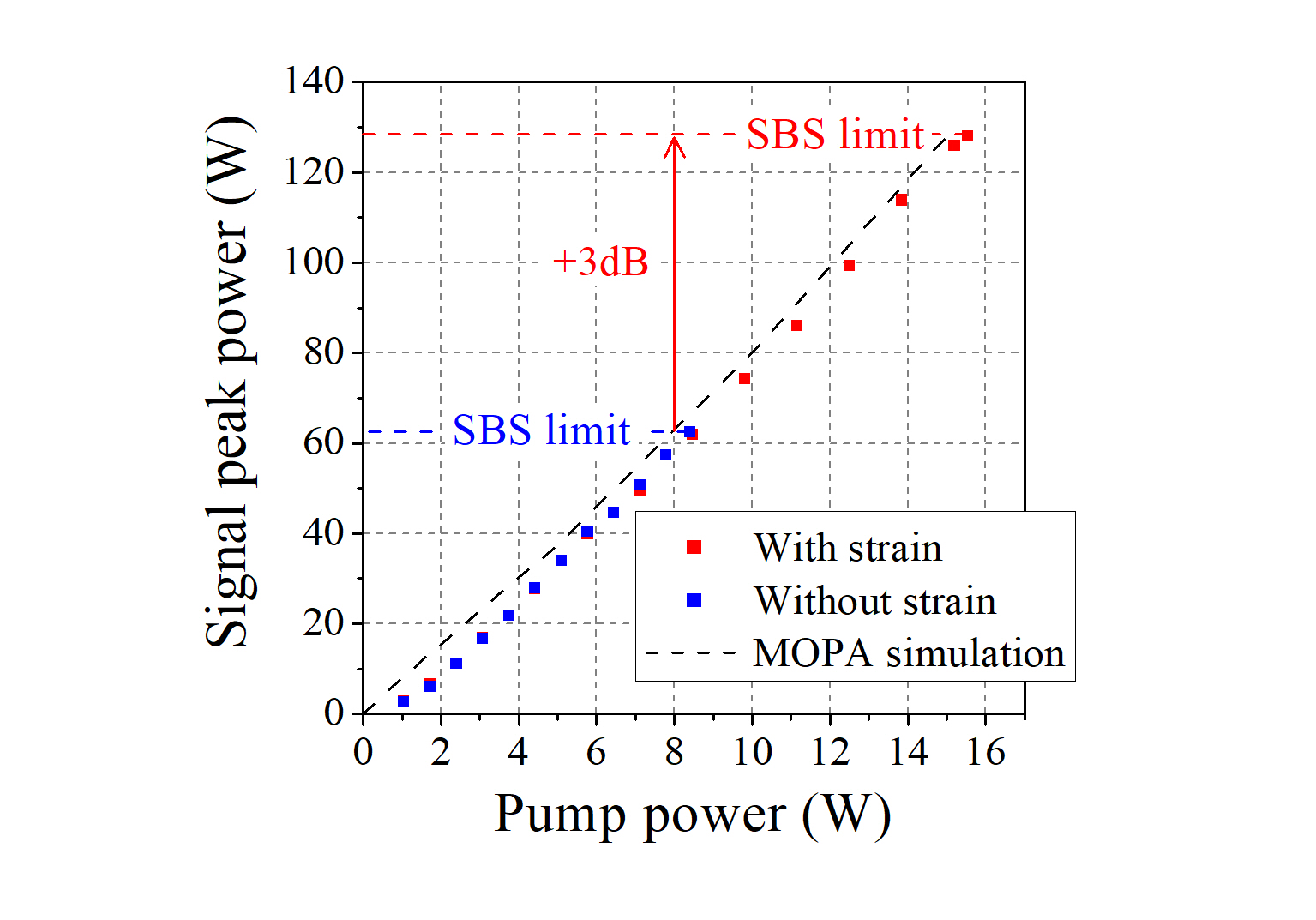}
\caption{Measurements of the signal peak power at the output of the fiber amplifier as a function of the pump power, before (blue) and after (red) applying the strain gradient. The SBS peak power threshold increases from 62~W to 128~W after applying the compressive strain. The dashed curve corresponds to the numerical simulation used to infer the laser gain profile $g_a(z)$ along the fiber.}
\label{fig_MOPA_SBS_treshold}
\end{figure}

As described above, the laser seed delivers single-frequency Gaussian pulses with FWHM $\tau_p=400$~ns and a repetition rate $f_\text{rep}=30$~kHz. The average signal power $\left\langle P_s\right\rangle$ is converted into pulse peak power $P_p= 0.94\times \left\langle P_s\right\rangle /(\tau_p.f_\text{rep})$, knowing that the inter-pulse amplified fluorescence is negligible for this repetition rate. 
The resulting measurements of the signal peak power as a function of the pump power is presented on Figure \ref{fig_MOPA_SBS_treshold}. Without strain on the fiber, the maximum peak power is limited to 62~W by SBS, whereas the critical power is raised to 128~W once we applied the compressive gradient. The peak power thus increased by a factor $\text{PIF}=2.1$. Again this value is close to the attenuation $a_\text{meas}$ of the Stokes spectrum peak after compression, but the relation between these two quantities is nothing but trivial (see section \ref{sec_PIF}). 

With our compression gradient method, we demonstrate here an improvement of about 3~dB for the SBS threshold power, for a maximum compressive strain close to 0.6\%. We did manage to reach higher compression, up to 1.5\%, and higher SBS threshold improvement, up to 6~dB, but at the expense of the core birefringence of the PM fibers. In such fibers, a transverse anisotropic stress is applied to separate the effective refractive indices of orthogonally polarized optical eigenmodes, using elliptic cores or stress rods. Experimentally, we found that a strong longitudinal strain would degrade the polarization extinction ratio (PER), most probably because it alters the transverse stress. From preliminary tests, we estimate that applying our compression method on active fibers, where the birefringence is generally lower than in passive fibers, a PER higher than 16~dB (typically required for coherent lidar systems) is compatible with PIF values of about 4-5~dB.
This value is comparable with the results obtained with other SBS attenuation methods that are compatible with single-frequency fiber amplifiers : about 3~dB for two-tone amplification \cite{Dajani_2010}, 3~dB for homogeneous broadening \cite{Le_Gouet_2019}, 3-4~dB with thermal gradient \cite{Lou_2020}, and up to 6~dB for specific core compositions \cite{Nakanishi_2006, Cavillon_2018}.




\section{Conclusion}
In order to raise the peak power that can be delivered by a passive fiber or by a fiber amplifier, one well-known and efficient method consists in applying a strain gradient along the fiber to create an inhomogeneous broadening of the SBS gain. In general, the strain gradient is obtained by pulling on the fiber, which can affect its lifetime. Here we detail our method to apply a compressive strain gradient on uncoiled fibers. We also develop our understanding of the strain gradient effect on both the SBS Stokes spectrum and the SBS threshold power. The good agreement between our calculations and measurements illustrates the relevance of the former.

The very next step will consist in using the calculations derived here to predict the strain effect on the SBS threshold, in the case of a MOPA. Both laser signal and SBS Stokes field varying along the fiber, the definition and calculation of the SBS threshold are also modified. This configuration will be studied in a future paper, together with a method to determine the optimal strain gradient.

In the meantime, we started to apply our method to strain efficiently a large mode area (LMA) fiber, with a larger clad diameter. The challenge here is that for a required strain, the force to apply scales quadratically with the fiber diameter, but preliminary tests on a passive fiber already show promising results.



\section*{Disclosures}

The authors declare no conflicts of interest.


\bigskip

\bibliography{2020_compressed_amp_bib}

\end{document}